\documentclass[a4paper,fleqn,usenatbib,authoryear,square]{mnras}

\usepackage{lipsum}
\usepackage{newtxtext,newtxmath}
\usepackage[T1]{fontenc}
\usepackage{ae,aecompl}
\usepackage{hyperref}
\usepackage{xcolor}

\usepackage{graphicx}	% Including figure files
\usepackage{amsmath}	% Advanced maths commands
\usepackage{amssymb}	% Extra maths symbols

\title[Flaring chromosphere oscillations]{The effect of a solar flare on chromospheric oscillations}

\author[D.C.L. Millar]{
David C. L. Millar,$^{1}$\thanks{E-mail: d.millar.2@research.gla.ac.uk}
Lyndsay Fletcher,$^{1,2}$
Ryan O. Milligan$^{1,3}$
\\
$^{1}$School of Physics \& Astronomy, University of Glasgow, Glasgow G12 8QQ, UK\\
$^{2}$Rosseland Centre for Solar Physics, University of Oslo, P.O.Box 1029 Blindern, NO-0315 Oslo, Norway\\
$^{3}$Astrophysics Research Centre, School of Mathematics and Physics, Queen's University Belfast, BT7 1NN, Northern Ireland, UK}

\date{Accepted XXX. Received YYY; in original form ZZZ}

\pubyear{2020}

\hypersetup{draft}

\begin{document}
\label{firstpage}
\pagerange{\pageref{firstpage}--\pageref{lastpage}}
\maketitle

%\begin{abstract}
%Oscillations in the chromosphere have been observed in quiet Sun conditions for some time, and more recently during solar flares and other energetic events. We aim to investigate the effect of a solar flare on the oscillations which are ubiquitous in the chromosphere: the 3-minute oscillations the period of which is governed by the chromospheric acoustic cut-off frequency. Using imaging spectroscopy data from the CRISP instrument, we searched for signals of oscillatory behaviour in timeseries drawn from small groups of pixels both before and after an M-class flare. Using several wavelengths from across the Calcium \textsc{ii} 8542 and H$\alpha$ spectral lines, we studied a range of heights in the solar atmosphere. We found that the oscillatory signals from the chromosphere were altered during the flare activity in two ways: the location of the oscillatory signals had changed and the lengths of the oscillations had tended to increase. Both of these results can be  explained by the restructuring of the magnetic fields in the chromosphere during the flare activity, and this is backed up by images of coronal loops in this active region showing clear changes in magnetic connectivity. A change in the atmosphere's magnetic environment can affect its acoustic cut-off frequency, and hence the periods of the waves that can propagate into the chromosphere.
%\end{abstract}

% new abstract

\begin{abstract}
    Oscillations in the solar atmosphere have long been observed in quiet conditions, and increasingly also in data taken during solar flares. The chromosphere is known for its 3-minute signals, which are particularly strong over sunspot umbrae. These signals are thought to be driven by photospheric disturbances and their periods determined by the chromosphere's acoustic cut-off frequency. A small number of observations have shown the chromospheric 3-minute signals to be affected by energetic events such as solar flares, however the link between flare activity and these oscillatory signals remains unclear. In this work we present evidence of changes to the oscillatory structure of the chromosphere over a sunspot which occurs during the impulsive phase of an M1 flare. Using imaging data from the CRISP instrument across the H$\alpha$ and Ca \textsc{ii} 8542 \AA\thinspace spectral lines, we employed a method of fitting models to power spectra to produce maps of areas where there is evidence of oscillatory signals above a red noise background. Comparing results taken before and after the impulsive phase of the flare, we found that the oscillatory signals taken after the start of the flare differ in two ways: the locations of oscillatory signals had changed and the typical periods of the oscillations had tended to increase (in some cases increasing from <100s to $\sim$200s). Both of these results can be explained by a restructuring of the magnetic field in the chromosphere during the flare activity, which is backed up by images of coronal loops showing clear changes to magnetic connectivity. These results represent one of the many ways that active regions can be affected by solar flare events.
    
\end{abstract}


%%%%%%%%%%%%%%%%%%%%%%%%%%%%%%%%%%%%%%%%%%%%%%%%%%

%%%%%%%%%%%%%%%%% BODY OF PAPER %%%%%%%%%%%%%%%%%%

\section{Introduction}\label{sec:intro}

Much of the variation we see in our observations of the Sun is essentially random in nature, and is described as noise. However, there are many sources of true periodic signals which can be identified, such as the 11-year sunspot cycle, the \emph{p}-modes seen at the photosphere and the 3-minute chromospheric oscillations. These periodicities are well established and are always present in the Sun, but there are also many transient phenomena which produce oscillatory signals. Examples include coronal loop oscillations \citep{aschwanden_1999_loop-oscillations, loop_osc_review} and quasi-periodic pulsations (QPPs) seen during solar flare activity, which have been observed across the electromagnetic spectrum and at timescales ranging from sub-second to hours \citep{Van-Doorsselaere2016}.

The 3-minute chromospheric observations were first reported in the 1970s \citep{bhatnagar_1972} and their origin is thought to be linked to the photospheric \emph{p}-modes, driven by activity in the solar interior. Different explanations have been suggested and modeled in the past, such as a resonant chromospheric cavity \citep{Leibacher_stein_1981}, but the prevailing theory is that the 3-minute signature is an intrinsic property of the chromosphere caused by its acoustic cut-off frequency \citep{fleck_schmitz_1991}. The cut-off frequency in an isothermal atmosphere is given by:
\begin{equation}
    \label{eq:cut-off}
    \omega_c = \frac{\gamma g}{2c_S} = \sqrt{\frac{\gamma \mu g^2}{4RT}},
\end{equation}
where $\gamma$ is the adiabatic index, $\mu$ is the mean molecular mass, $g$ is the gravitational acceleration, $R$ is the gas constant and $T$ is the temperature. This property puts a lower limit on the frequency of acoustic (pressure) waves which can propagate through a medium, filtering out disturbances at $\omega < \omega_c $ from the photosphere below \citep{Lamb_1909}. Using typical values for the chromosphere yields $\omega_c\approx 0.03\ \rm{rad \ s^{-1}}$, equivalent to a period of approximately 200 seconds.

However, in the magnetised solar atmosphere it is necessary to think beyond the acoustic modes. In regions of high magnetic field, photospheric disturbances can convert into magnetoacoustic waves in regions where the Alfv\'{e}n speed and sound speed are similar, and the resulting waves travel along magnetic field lines. Due to the Sun's gravitational field, the waves become magnetoacoustic gravity waves \citep[MAG waves:][]{Bel_Leroy_1977A&A....55..239B} and the cut-off frequency depends on the angle between the magnetic and gravitational fields. This leads to strong 3-minute signatures above sunspot umbrae in the chromosphere, and also to running penumbral waves \citep[RPWs:][]{Jess_2013ApJ...779..168J} -- where the period of oscillatory signals increases as one moves radially out from the centre of a sunspot, as the magnetic field inclination to vertical increases \citep{Reznikova_2012_3mins,Sych_2020_ApJ...888...84S}.

There have been studies which suggest that activity from higher up in the atmosphere can affect the ubiquitous 3-minute oscillations, for example plasma downflows \citep{Kwak} and solar flares \citep{milligan_2017,Kosovichev_2007ApJ...670L.147K} exciting the chromosphere, and causing enhanced signals at the resonant period. Extremely powerful flares can cause sunquakes, a chromospheric signature of which has recently been observed by \cite{Quinn_2019}. Specifically, oscillations over sunspots in flaring regions have been observed in the past \citep{Kosovichev_2007ApJ...670L.147K,sych_2009}, but as of now the link between flare activity and sunspot oscillations is unclear.

In this paper, we study the effect of flare activity on the oscillatory signals present in an active region above a sunspot, focusing particularly on the umbra/penumbra and flare footpoints, with imaging spectroscopy data from the CRISP instrument at the Swedish Solar Telescope. We aimed to identify oscillatory signals present both before and after the onset of the flare, to observe if the flare activity had induced oscillatory behaviour, or affected the signals which were present beforehand.

In Section~\ref{sect:event_data} we overview the flare, describe the datasets used and the initial processing steps; in Section~\ref{sec:spectral_analysis} we outline the methods used to identify oscillatoy signals in the data; Section~\ref{sec:results} outlines the main results from our analysis; we discuss limitations to our methods and possible interpretations of the results in Section~\ref{sec:discussion} before concluding in Section~\ref{sec:conclusions}. 

\section{Event and datasets}\label{sect:event_data}
\begin{figure}
	\includegraphics[width=\columnwidth]{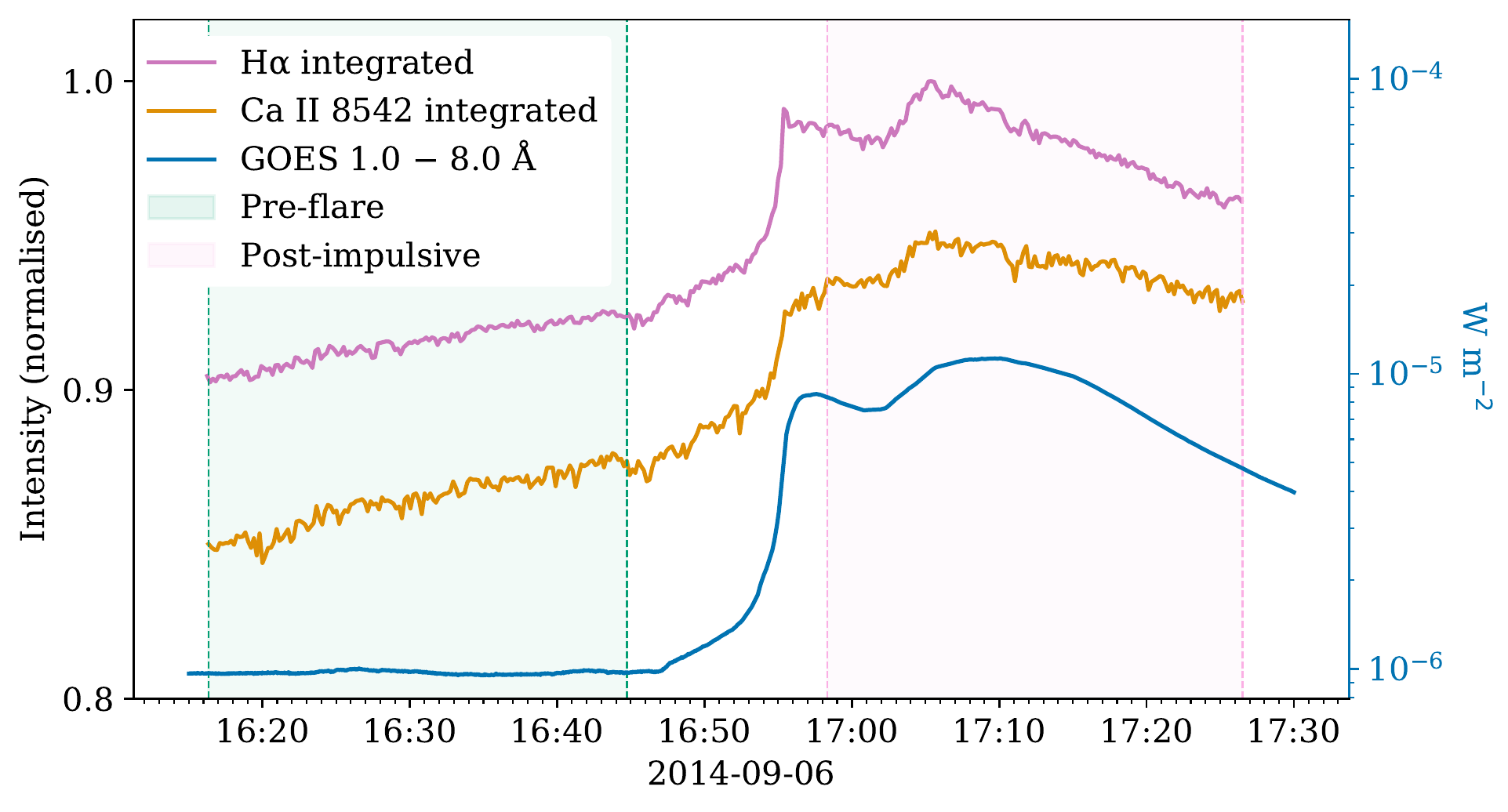}
    \caption{Lightcurves of GOES 1-8 \AA\ and the H$\alpha$ and Ca\thinspace\textsc{ii}~8542 spectral lines from CRISP integrated across the field of view. The shaded areas indicate the two $\sim$30 minute periods which are analysed separately: ``pre-flare'' (green) and ``post-impulsive'' (pink).}
    \label{fig:goes}
\end{figure}

\begin{figure}
    \centering
	\includegraphics[width=\columnwidth]{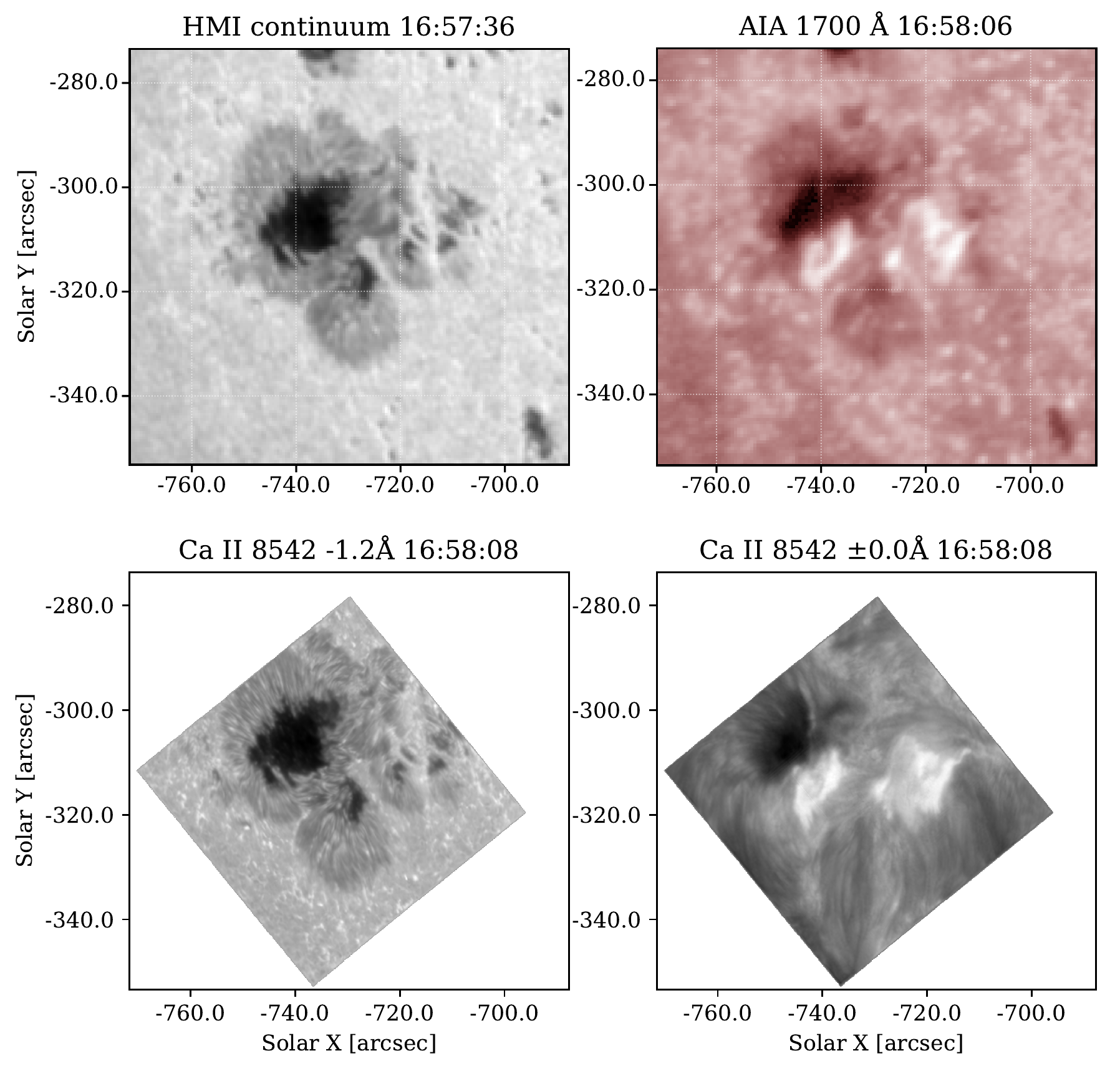}
    \caption{The area observed by CRISP at a time just after the flare onset, in several wavelengths from CRISP and SDO. The line wing (lower-left) displays the photosphere, similar to the HMI/SDO continuum image (upper-left), while the line core (lower-right) shows the chromosphere and can be compared to the AIA/SDO 1700 \AA\ image (upper-right).}
    \label{fig:context}
\end{figure}

The M1.1 flare SOL2014-09-06T17:09 occurred in active region AR 12157 (-732", -302"). The analysis presented below is based on data from the CRisp Imaging SpectroPolarimeter \cite[CRISP:][]{2008ApJ...689L..69S} instrument at the Swedish Solar Telescope \cite[SST:][]{scharmer_2003}, and the Atmospheric Imaging Assembly \citep[AIA:][]{Lemen_2012_AIA} on the Solar Dynamics Observatory \cite[SDO:][]{Pesnell_2012_SDO}. The flare timeline is shown in Figure~\ref{fig:goes} with lightcurves from GOES 1-8\thinspace\AA\ and also from the CRISP instrument. Context images of the region during the flare activity from CRISP and SDO are shown in Figure~\ref{fig:context}, and the image coordinates used in the figures throughout this paper are displayed in relation to these images.
%https://www.overleaf.com/project/5bb6079d529d1133221cc28f

Individual pixels in the CRISP data show an extremely rapid increase in brightness during the flare onset at approximately 16:56 (see figure~\ref{fig:goes}), and the AIA channels contain many saturated pixels at this time. These two effects are detrimental to the analysis of oscillatory signals, and so two $\sim$30 minute periods were analysed, before the flare onset (16:15-16:45, ``pre-flare'') and after the initial brightening (16:57-17:27, ``post-impulsive'').

\subsection{CRISP}

The active region was observed by the CRISP instrument on the SST between approximately 15:30 and 17:30 in H$\alpha$ (6563\AA) and Ca \textsc{ii} 8542\AA, sampling the chromosphere and upper photosphere at 0.057\thinspace'' per pixel. The spectral dimension in H$\alpha$ covers $\pm$1.4\AA\ with a step size of 0.2\AA\ and Ca \textsc{ii} 8542 covers $\pm$1.2\thinspace \AA\ with step size 0.1\AA. Full spectral scans were obtained approximately every 11.6~seconds. The data are available on the F-CHROMA flare database (\texttt{f-chroma.org}).

The CRISP images are aligned such that solar North points upwards along the y-axis, and the field of view rotates and moves, keeping features of interest fixed at the same pixels throughout the observation. For each 30 minute analysis window (pre-flare and post-impulsive), only pixels which were in the field of view for the whole window were analysed. This accounts for the odd shape of the analysed field which will be seen in future figures. %\lf{(This accounts for the odd shape of the analysed field which will be seen in future images.)}

The images were originally of dimension (1398, 1473), but were re-binned by a factor of 10 to be dimension (139, 147), after clipping the edges of the images. Timeseries for analysis will be drawn from each 10x10 macropixel. This was done to reduce the effects of seeing, and to reduce the required computation time.

\subsection{AIA}
The Atmospheric Imaging Assembly (AIA) on NASA's Solar Dynamics Observatory (SDO) provides full disk images of the Sun in 8 EUV channels with a cadence of 12 seconds, and 2 UV channels with a cadence of 24 seconds, all at a resolution of 0.6\thinspace'' per pixel. The various channels observe a range of temperatures in the solar atmosphere, roughly corresponding to different heights, from photosphere to corona. The channels used in this analysis were 1600\AA\thinspace and 1700\AA\thinspace. Other AIA channels were not used, as they were greatly affected by the brightness increase caused by the flare activity, with much saturation and blooming continuing well after the flare peak, mostly in bright loop structures.

AIA images for the chosen channels were obtained and prepped using the \texttt{SunPy} package \citep{SUNPY}, before cutting out an area of interest slightly larger than the CRISP field of view with the \texttt{submap} method, and accounting for solar rotation with the \texttt{mapsequence\_solar\_derotate} function from \texttt{sunpy.physics.solar\_rotation}.

\section{Spectral analysis}\label{sec:spectral_analysis}
To identify and characterise periodic signals in the data, we calculate and analyse the power spectral density (PSD) of each macropixel's timeseries. The PSDs were then fitted with different models which describe different shapes of power spectra we expected to observe. This method was first used to identify QPPs in flare data by \cite{inglis_2015}, and the methods described below are adapted from this work as well as studies by \cite{Auchere_2016_TC-wavelet-criticism} and \cite{Battams_2019_EUV-powerlaws}.

The power law nature of the background noise means that identifying true oscillatory signals is not as easy as identifying the frequency bin in the power spectrum with the highest value of power. Methods exist to filter out long period signals in the timeseries in order to identify interesting signals; examples include analysing the time-derivative of the data \citep{simoes_hudson_fletcher_2015}, Fourier filtering \citep{milligan_2017} and box-car smoothing \citep{Dolla_2012_boxcar}. The advantage of the spectral fitting method used here is that the original data is altered as little as possible.

\subsection{Obtaining power spectral density}
The PSD of a timeseries gives the ``power'' of periodic signals as a function of frequency. Firstly, each timeseries from individual macropixels was normalised by subtracting its mean and dividing by its standard deviation. The next step was to apodize the timeseries (multiplying by a Hann window function). The window function reduces spectral noise by removing the discontinuity between the first and last entry in the timeseries. The PSD is then obtained by performing a fast Fourier transform, and taking the absolute value of the result.

\subsection{Spectrum models}\label{sec:spectrum-models}
When the PSD is plotted in log-space, the shape of the spectrum can be identified. A flat line corresponds to perfect ``white'' noise, whereas a sloped line represents ``red'' noise. Formally, the colour of noise is determined by the $\alpha$ parameter in the following equation, which gives the power, $P$, as a function of frequency, $f$:

\begin{equation}
    P(f) = Af^{-\alpha}.
    \label{eq:powerlaw}
\end{equation}

$A$ is a constant which affects the vertical offset of the sloped line. 

%Some authors have identified periodic signals by first removing long-period trends in the timeseries by a variety of ``smoothing'' methods, before applying some analysis on the altered timeseries. However, this type of manipulation can lead to false results, depending on the chosen filter frequency or size of smoothing window.

In a signal originating from purely coloured noise with a single spectral index, the spectrum would be described well by Equation~\ref{eq:powerlaw}. However, in reality the timeseries we observe can be better described by an altered version of this simple power law. In this work we examine three different models to describe the observed PSDs.
%describing the various forms observed PSDs can take. %LF This seemed to imply that these are the only forms it can take

The first model (M1: equation~\ref{eq:M1}) which is used for the spectra is a power law with an additional constant, $C$, to describe a white noise element in the data from photon counting (technically this is two power laws summed together, one of which has index $\alpha=0$):

\begin{equation}
    \mathrm{M1} = Af^{-\alpha} + C.
    \label{eq:M1}
\end{equation}

\begin{figure}
    \centering
	\includegraphics[width=\columnwidth]{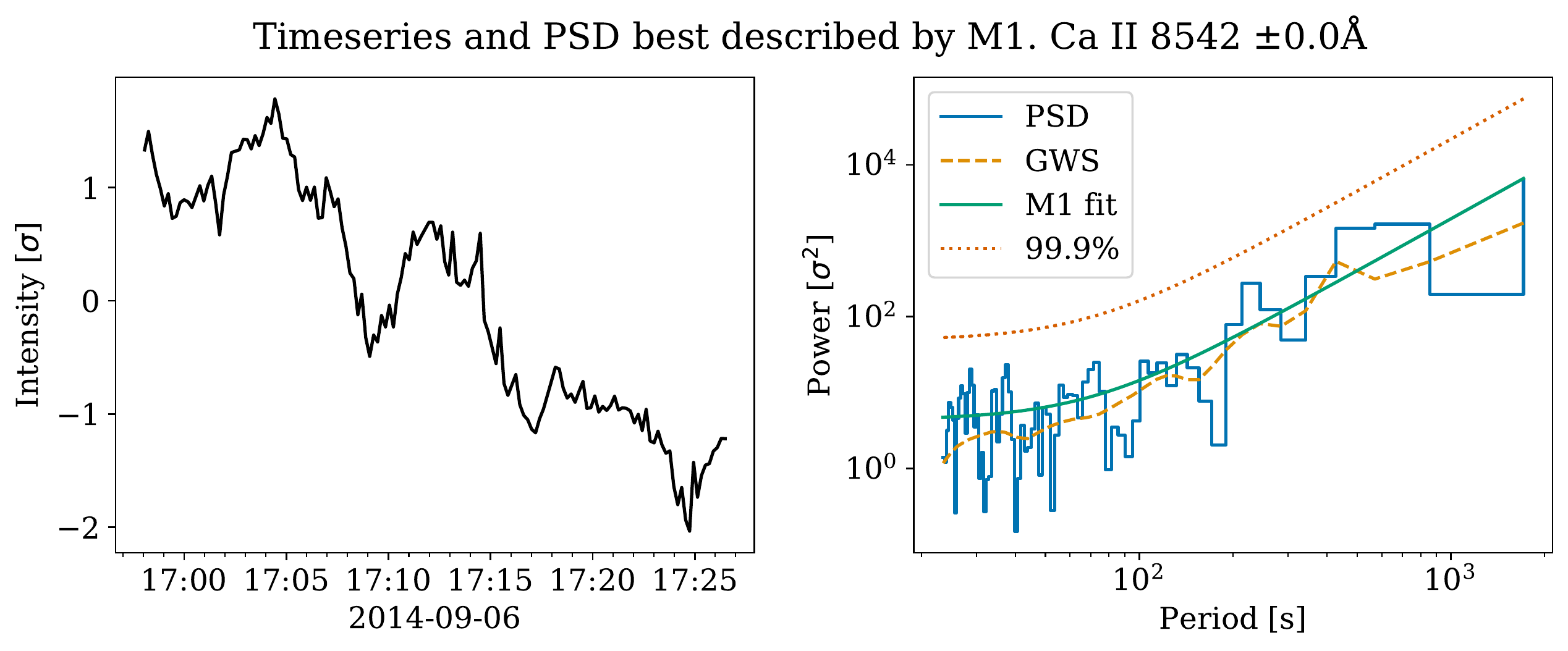}
    \caption{The timeseries (LH panel) and corresponding power spectral density (PSD, RH panel) taken from a macropixel from the line core of Ca \textsc{ii}~8542 at a location far from the main sunspot and the flare activity. Plotted with the PSD (blue stepped line) is the global wavelet spectrum (yellow dashed line), the mean spectrum fit by M1 (green line) and the 99.9\% significance level (orange dotted line).}
    \label{fig:rednoise}
\end{figure}

In Figure~\ref{fig:rednoise} a spectrum is shown which is best fit by the power-law noise model (M1) and shows the types of random variations which can be attributed to coloured noise. 
%Bins of power above this line have a <0.01\% confidence level of being attributed to the mean spectrum. The significance level is found by multiplying the mean spectrum by a factor $m$ determined by equation~\ref{eq:siglvl}
The second model (M2: equation~\ref{eq:M2}) includes a Gaussian bump term -- utilised also by \cite{inglis_2015} --  which is used to indicate enhanced oscillatory power in addition to the power described by M1. An additional three parameters are introduced to describe the height ($B_G$), width ($\sigma$) and position of the peak in frequency space ($\beta$):

\begin{equation}
    \mathrm{M2} = \mathrm{M1} + B_G\exp{\left(\frac{-(\ln{f}-\beta)^2}{2\sigma^2}\right)}.
    \label{eq:M2}
\end{equation}

\begin{figure}
    \centering
	\includegraphics[width=\columnwidth]{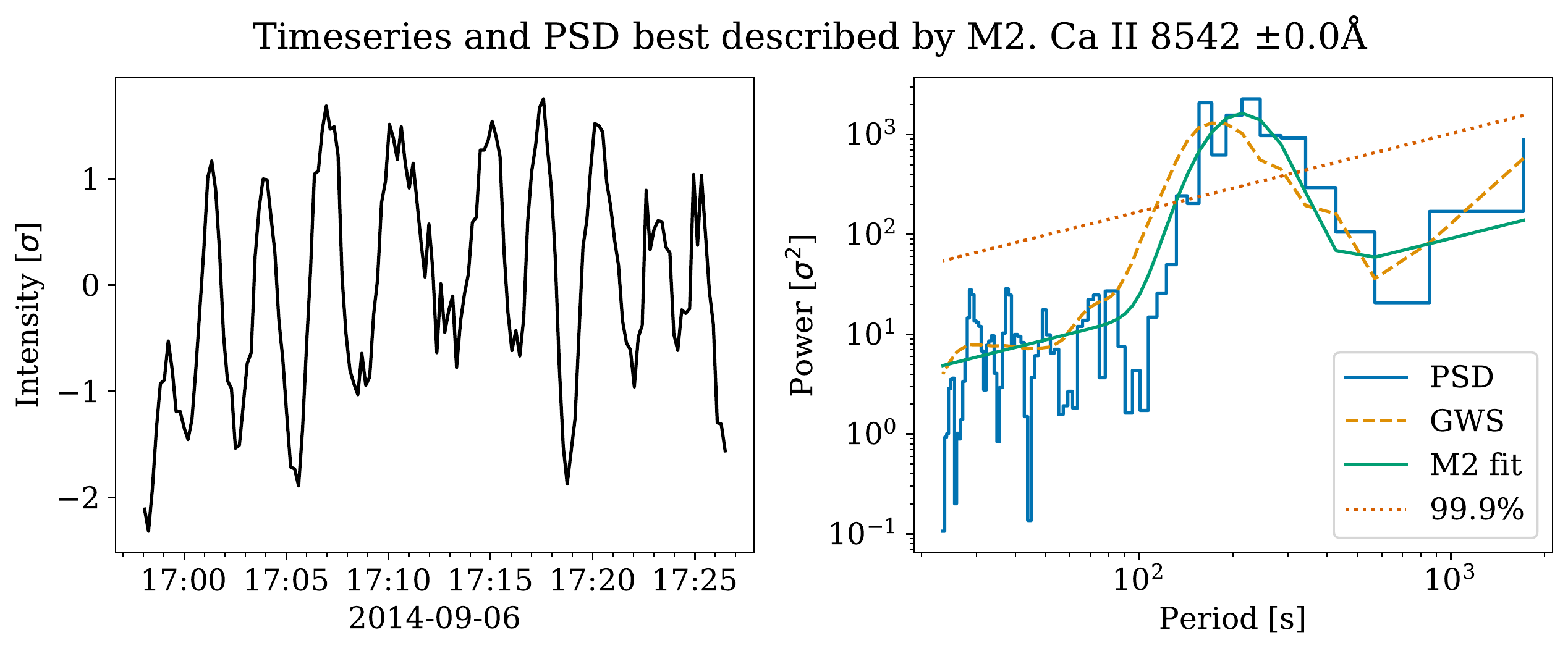}
    \caption{As in figure~\ref{fig:rednoise}, but showing results from a macropixel over the sunspot umbra. The confidence threshold (orange dotted line) is shown corresponding to the noise background if there was no bump present. The peak of the Gaussian bump occurs at $\beta=-5.36$ which corresponds to a period of $\approx210\ \rm{s}$.}
    \label{fig:gauss}
\end{figure}

An example of an M2 spectrum is displayed in Figure~\ref{fig:gauss}. The spectrum shown in this figure is clearly of a different nature to its Figure~\ref{fig:rednoise} counterpart, and it is well described by a bump rising above a coloured noise background. The timeseries has a clearly visible oscillatory signal which produces a PSD with a clear increase of power around the approximate period of the oscillations. The oscillations are not all of exactly equal length, and so the signal in the PSD is spread out somewhat, producing the bump. The parameter $\beta$ gives us an idea of the most common periodicities observed in the timeseries, and can be converted to a period via $P = e^{-\beta}$.

When observing some of the PSD fits using M1 and M2, it was found that often the PSD would appear to level off below certain frequencies, and this behaviour could not be described accurately by M1 or M2. We introduce a third model (M3: equation~\ref{eq:M3}) which utilises a kappa function, and has been used in the past by \cite{Auchere_2016_TC-wavelet-criticism} and \cite{Threlfall_2017}:

\begin{equation}
    \mathrm{M3} = \mathrm{M1} + B_K\left(1 + \frac{\nu^2}{\kappa\rho^2}\right)^{-\frac{\kappa+1}{2}}.
    \label{eq:M3}
\end{equation}

\begin{figure}
    \centering
	\includegraphics[width=\columnwidth]{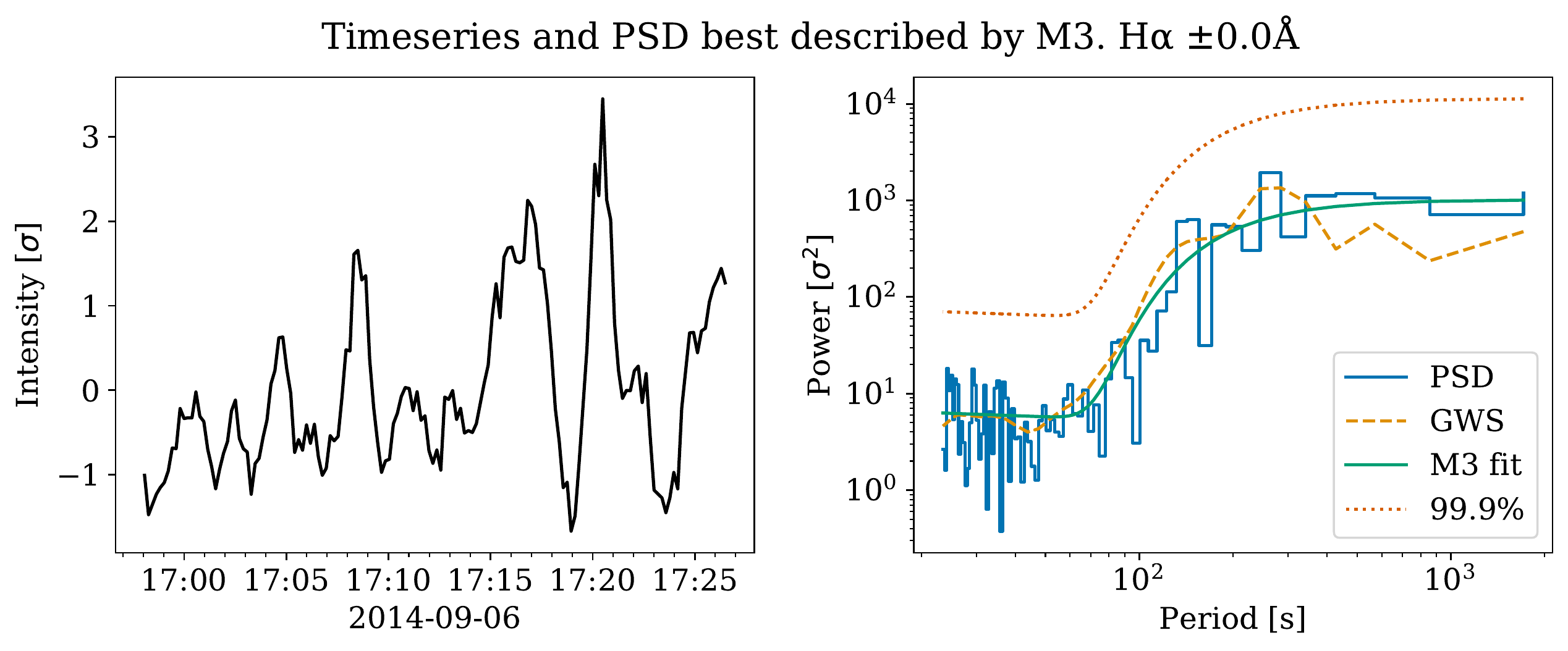}
    \caption{Similar to figures~\ref{fig:rednoise} and \ref{fig:gauss}, but for a macropixel from the core of H$\alpha$ over the sunspot penumbra, which has been best fit by the kappa function (M3).}
    \label{fig:kappa}
\end{figure}

In the above equation, $\rho$ describes the width of the kappa function, $\kappa$ describes its extent into the high-frequency wing, and $B_K$ gives its height. See Figure~\ref{fig:kappa} for an example of a spectrum which was fitted by M3. It is clear from the timeseries that there is oscillatory behaviour here, however it is not sufficiently well localised to any particular frequency to create a defined bump. While it is possible that the kappa function gives some information about oscillatory behaviour it is more complex to define than the Gaussian bump case.  In Figure~\ref{fig:kappa} the oscillations are drifting in period over the $\sim$30 minute window, which could create a wide increase in power which manifests in the kappa shaped bulge. An apparent flattening of the spectrum could be caused by enhanced power at shorter periods or a decrease in power at longer periods.

While there have been theories put forward as to why a power law noise spectrum describes a solar timeseries -- for example due to many small energy deposition events \citep{Aschwanden2011}, there is little explanation for why a PSD should take the shape of a Gaussian bump or kappa function, as opposed to any other shape. \cite{Battams_2019_EUV-powerlaws} used a Lorentzian bump instead of Gaussian, motivated by its physical meaning in the context of a damped oscillator. However we found more success fitting the PSDs from this dataset when using the Gaussian option. In this context, these models can be seen as tools to identify PSDs which deviate from simple coloured noise, rather than descriptors of the physical processes behind the observed timeseries and PSDs.

The fitting process was performed in Python using the \texttt{curve\_fit} function from the \texttt{scipy} library \citep{SCIPY}. The individual data points for the power spectra span many orders of magnitude, so the squared residuals must be weighted to be comparable. Following \cite{Auchere_2016_TC-wavelet-criticism}, the global wavelet spectrum (GWS) was used as a weighting function. The GWS is a proxy for the PSD of the timeseries, and is calculated by time-averaging the wavelet transform, as described by \cite{torrence_compo_1998}. It is assigned to the ``sigma'' argument in \texttt{curve\_fit}.

% Here's what I found from Auchere's IDL code he released beside his 2016 paper:
% The power spectrum is fitted using curvefit. In order for the least squares fit to converge, the
% residuals (differences between data points and fit) must be comparable at each frequency,
% This is achieved by normalizing the squared residuals by the variance of the data.
% At each frequency nu, the Fourier spectrum is distributed as 0.5sigma(nu)chi2 (degree 2 chi2, Eq. 11).
% Its standard deviation is thus (0.5sigma(nu))2=sigma(nu). The standard deviation of the power
% spectrum is thus equal to its mean at each frequency. Therefore we need to weight the fit by the result of the fit squared (the mean power squared as a function of frequency) ... But that is OK because only a first order estimate of the mean power is needed. We can simply use the time-averaged wavelet spectrum as a proxy.

\subsection{Identification of significant oscillatory signals}

Identifying true oscillatory signals above noise is a very difficult task without visually inspecting each timeseries and spectrum, and therefore we set a number of criteria that each spectral fit must first meet. The first criterion to identify an M2 fit is that the Gaussian bump model must describe the observed data better than M1 and M3, based on a weighted residuals squared (WRS) measurement of each fit (weighted using the global wavelet spectrum). M2 and M3 have three more free parameters than M1, and so these would be expected to fit better in most circumstances. An F-test is used when comparing models with different numbers of parameters, with the F-statistic defined as:

\begin{equation}
    F = \frac{\left( \frac{WRS_1-WRS_2}{p_2-p_1}\right)}{\left(\frac{WRS_2}{n-p_2}\right)}.
    \label{eq:F-test}
\end{equation}

% \begin{figure}[hbtp]
%     \centering
%     \includegraphics[width=\linewidth]{spectral-fit-example.pdf}
%     \caption{The left hand panel shows the timeseries, normalised by subtracting the mean and dividing by the standard deviation. The PSD is shown in the right panel, along with the fits to models 1 and 2 which were output by the fitting function, the weighting function and the 95\% significance level based on the background spectrum. The M2 parameters determined for this pixel are: $A\approx0.006$, $B\approx679$, $C\approx3.6$, $\alpha\approx1.68$, $\sigma\approx0.16$, $\beta\approx-4.97$ (bump located at $e^{4.97}\approx144$ seconds). }
%     \label{fig:spectral-fit-example}
% \end{figure}

The null hypothesis for the F-test is that M1 describes the data as well or better than M2 (or M3) and a $p$-value to reject it is obtained based on the value of $F$ in the $F(p_2-p_1, n-p_2)$ distribution. This $p$-value threshold is $p<0.001$ for the majority of the analysis. It should be noted that a $p$-value below the threshold does not indicate that M2 (or M3) describes the data perfectly (or well at all), just that it is preferred over M1, and the test is expected to be wrong 0.1\% of the time.

Spectra which are best described by M2 and which pass this F-test were then identified to have two main components: a Gaussian bump and a background noise spectrum described by M1. The background spectrum is found by setting the $B$ parameter to zero, removing the bump. The significance of the bump was tested against the background noise level, based on a confidence threshold obtained using the parameter \textit{m} from the following equation:

\begin{equation}
    m = -\ln (1-X^{1/N}).
    \label{eq:siglvl}
\end{equation}

This equation states that each of the $N$ frequency bins of the PSD has a probability $X$ of being $m$ times greater than the background noise spectrum. It is obtained from the fact that the Fourier spectrum is $\chi^2_2$ distributed around the mean power value at each frequency, and is explained fully in \cite{Auchere_2016_TC-wavelet-criticism}. The spectral bump is taken to be significant if the fitted spectrum is more than $m$ times the background spectrum at a confidence level of 99.9\%. By applying these two different significance tests to the data the likelihood of identifying a timeseries which does not contain a significant deviation from the standard background noise was greatly reduced.

The final step in determining the significance of each Gaussian bump, was to remove results which contained bumps very close to the edges of the spectra. Results were only kept which satisfied $-6.1 < \beta < -3.91$, these limits corresponding to approximately 450 and 50 second periods, respectively. 

\section{Results}\label{sec:results}

\begin{figure*}
    \centering
	\includegraphics[width=2\columnwidth]{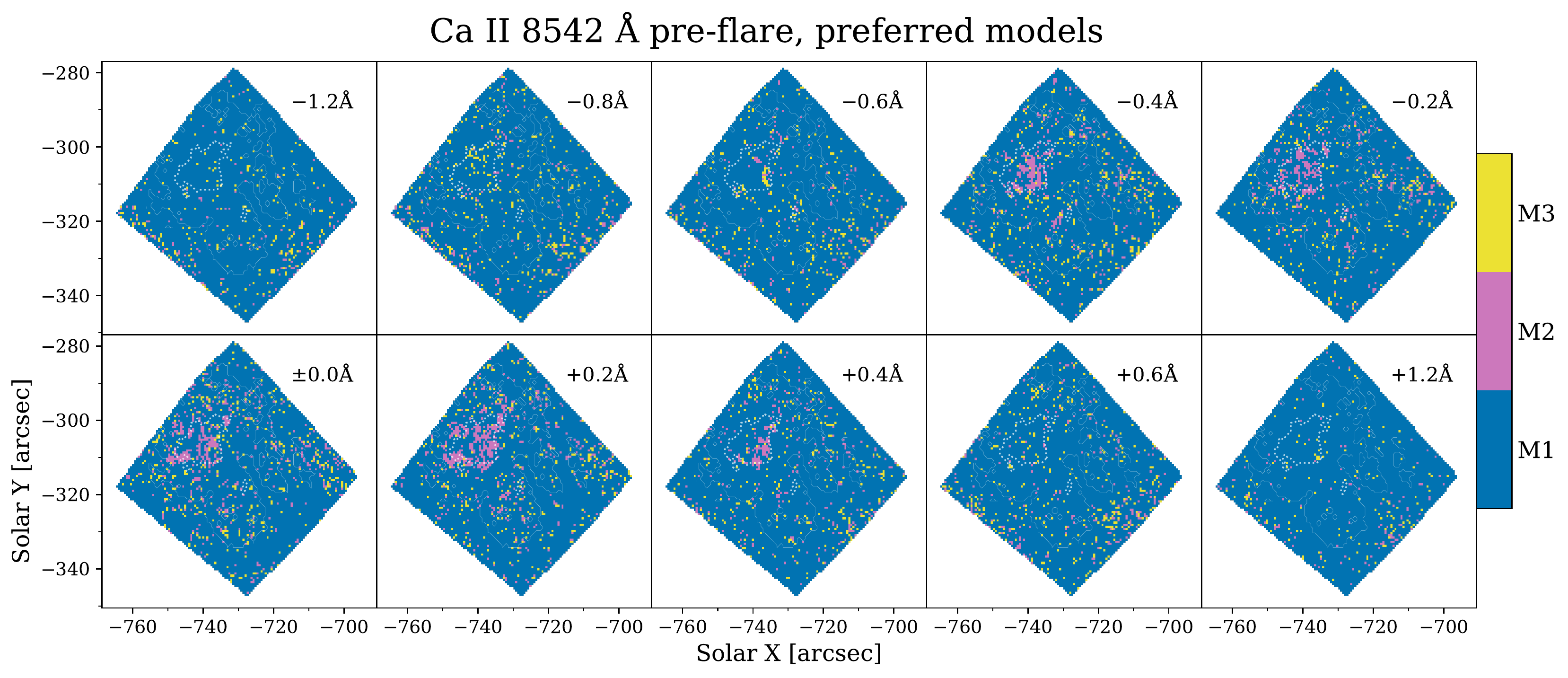}
    \caption{The models which best describe the spectra of individual macropixels at several points across the Ca \textsc{ii} 8542\AA\thinspace line, during the pre-flare period (16:15-16:45). Each model is assigned a different colour, and the edges of the sunspot umbra and penumbra (drawn from 40\% and 75\% intensity levels of the first wing image) are shown in dotted and solid contours, respectively.}
    \label{fig:preferred_8542_pre}
\end{figure*}

\begin{figure*}
    \centering
	\includegraphics[width=2\columnwidth]{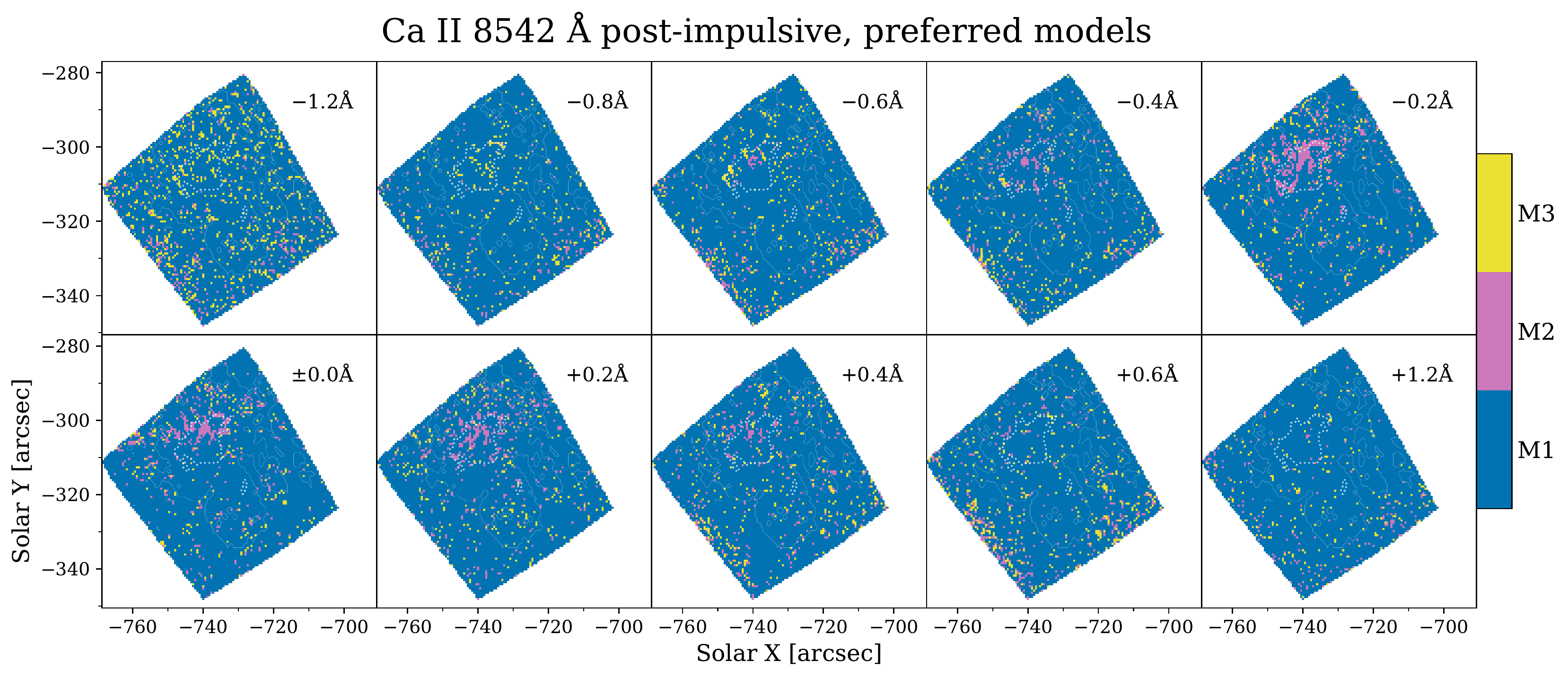}
    \caption{Similar to Figure~\ref{fig:preferred_8542_pre} but during the post-impulsive period (16:57-17:27). The shape of the plots is different than in the pre-flare case because of the rotation of the CRISP field of view. Note the northern umbra/penumbra border with concentrations of M2 fits at several wavelengths.}
    \label{fig:preferred_8542_post}
\end{figure*}

\begin{figure*}
    \centering
	\includegraphics[width=2\columnwidth]{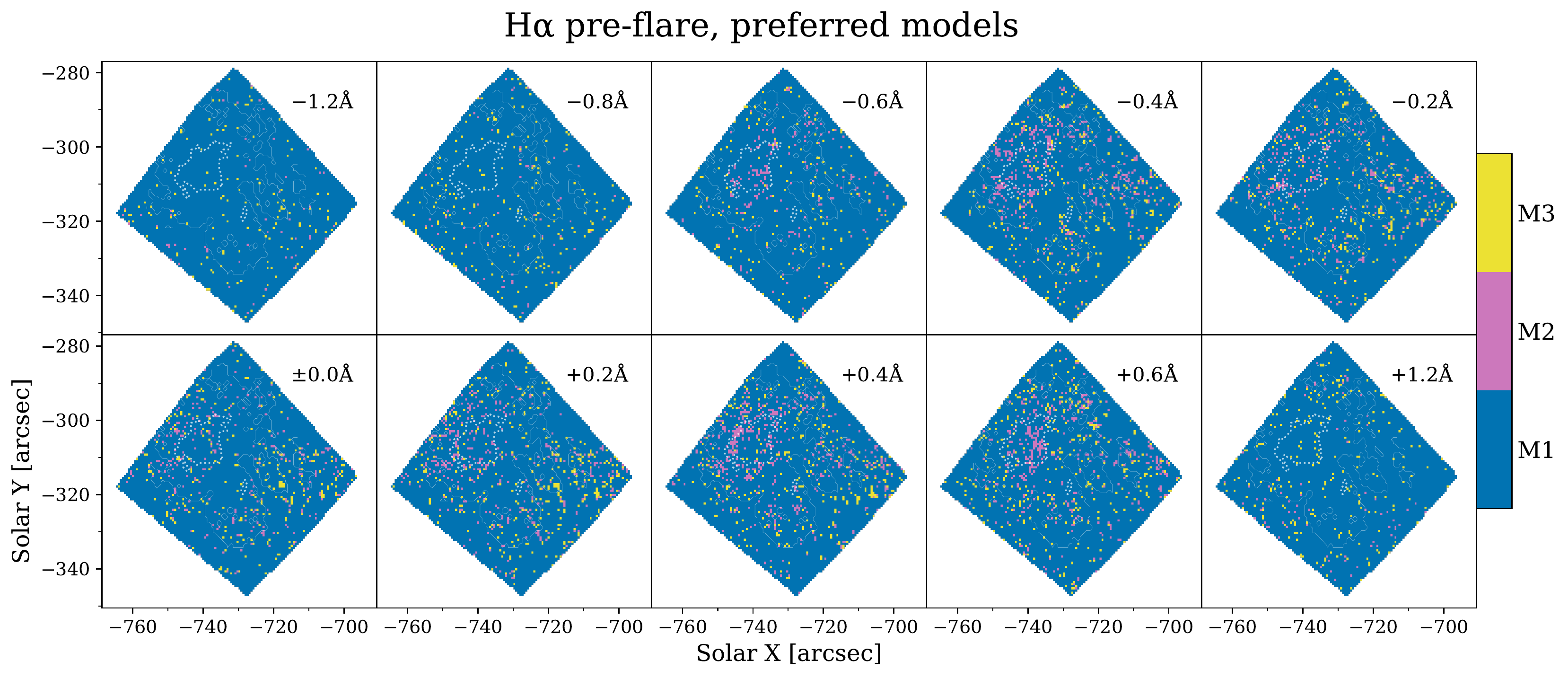}
    \caption{Similar to Figure~\ref{fig:preferred_8542_pre} but for the H$\alpha$ line.}
    \label{fig:preferred_Halpha_pre}
\end{figure*}

\begin{figure*}
    \centering
	\includegraphics[width=2\columnwidth]{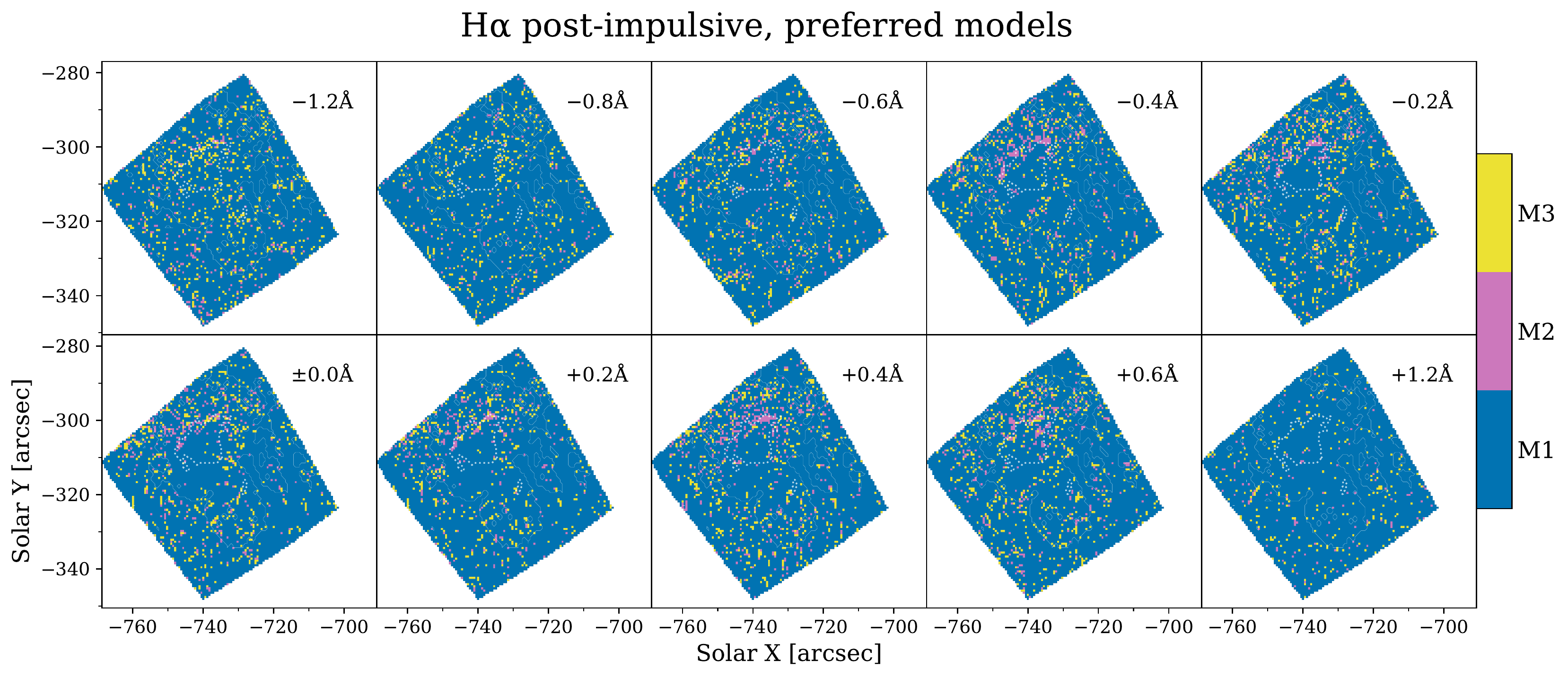}
    \caption{Similar to Figure~\ref{fig:preferred_8542_pre} but for H$\alpha$, after the initial flare activity.}
    \label{fig:preferred_Halpha_post}
\end{figure*}

\subsection{Preferred models}\label{sec:preferred}
The methods described in Section~\ref{sec:spectral_analysis} allowed a preferred model (M1, M2 or M3) to be assigned to each macropixel in our pre-flare and post-impulsive data, at several wavelength points across the H$\alpha$ and Ca \textsc{ii} 8542\AA\thinspace spectral lines. Visualisations are found in figures~\ref{fig:preferred_8542_pre}-\ref{fig:preferred_Halpha_post}, and display the areas of oscillatory signals at these different wavelength points, overplotted with contours showing the outlines of the sunspot umbra and penumbra. 

The first thing to note from these plots is that M1, the power-law coloured-noise model was the most common result, being the preferred result in 83\% of all pixels, compared with 6\% for M2 which includes the Gaussian bump, and 11\% for the M3 kappa function. This is unsurprising as M1 is the simplest and therefore most generally applicable shape of the three. Table~\ref{tab:percents} shows the percentage of M2 and M3 fits which were assigned to macropixels in select wavelengths and in both the pre-flare and post-impulsive time periods. 

Broadly, it appears that more M2 fits are found for power spectra at wavelengths closer to the centres of the lines, out to $\pm0.4 $\AA, and areas with many M2 fits are found over the sunspot umbra, and around its border. As we move in wavelength away from the line core towards the wings, the occurrence of M2 fits drops considerably. Many more pixels in the wings show fits to M3, especially in the post-impulsive data for both lines.

Although there appears to be correlation between the positions of M2 fits and the sunspot umbra, there is not much distinction between the penumbra of the sunspot and the surrounding area.

\begin{table}

\centering
\caption{Percentage of all macropixels which had preferred model fits M2 (Gaussian bump) and M3 (kappa function) for each wavelength and for the pre-flare and post-impulsive periods.}
\label{tab:percents}
\begin{tabular}{ccccc}
\hline
Wavelength & Pre M2 & Pre M3 & Post M2 & Post M3 \\
\hline

8542 $-1.2$\AA & 1.9\% & 2.9\% & 3.9\% & 9.5\% \\
%8542 $-1.1$\AA & 1.7\% & 3.0\% & 2.0\% & 2.2\% \\
%8542 $-1.0$\AA & 2.3\% & 3.8\% & 2.4\% & 2.8\% \\
%8542 $-0.9$\AA & 2.3\% & 3.7\% & 2.8\% & 3.2\% \\
8542 $-0.8$\AA & 3.1\% & 4.8\% & 2.5\% & 4.0\% \\
%8542 $-0.7$\AA & 2.6\% & 4.2\% & 3.2\% & 6.4\% \\
8542 $-0.6$\AA & 2.7\% & 3.9\% & 3.2\% & 4.6\% \\
%8542 $-0.5$\AA & 3.8\% & 4.3\% & 2.9\% & 2.9\% \\
8542 $-0.4$\AA & 5.7\% & 4.9\% & 4.2\% & 4.0\% \\
%8542 $-0.3$\AA & 5.2\% & 4.4\% & 6.4\% & 3.5\% \\
8542 $-0.2$\AA & 5.7\% & 3.8\% & 7.1\% & 3.3\% \\
%8542 $-0.1$\AA & 9.1\% & 5.9\% & 6.4\% & 3.0\% \\
8542 $\pm0.0$\AA & 7.1\% & 4.7\% & 5.6\% & 2.7\% \\
%8542 $+0.1$\AA & 6.9\% & 3.6\% & 4.6\% & 2.8\% \\
8542 $+0.2$\AA & 7.8\% & 4.1\% & 5.7\% & 3.1\% \\
%8542 $+0.3$\AA & 5.2\% & 3.2\% & 5.4\% & 3.5\% \\
8542 $+0.4$\AA & 4.1\% & 3.6\% & 4.2\% & 3.8\% \\
%8542 $+0.5$\AA & 3.4\% & 3.6\% & 4.4\% & 4.5\% \\
8542 $+0.6$\AA & 3.2\% & 4.4\% & 4.4\% & 4.6\% \\
%8542 $+0.7$\AA & 2.3\% & 3.0\% & 3.5\% & 4.6\% \\
%8542 $+0.8$\AA & 1.7\% & 2.8\% & 3.3\% & 3.9\% \\
%8542 $+0.9$\AA & 1.5\% & 2.3\% & 3.3\% & 4.0\% \\
%8542 $+1.0$\AA & 1.6\% & 2.0\% & 3.4\% & 4.0\% \\
%8542 $+1.1$\AA & 1.4\% & 2.5\% & 2.5\% & 2.6\% \\
8542 $+1.2$\AA & 1.6\% & 1.9\% & 2.2\% & 2.5\% \\
\hline
%H$\alpha -1.4$\AA & 0.7\% & 1.4\% & 2.3\% & 7.2\% \\
H$\alpha -1.2$\AA & 0.7\% & 1.7\% & 3.2\% & 8.1\% \\
%H$\alpha -1.0$\AA & 0.7\% & 1.6\% & 2.0\% & 4.0\% \\
H$\alpha -0.8$\AA & 0.9\% & 2.5\% & 2.2\% & 5.9\% \\
H$\alpha -0.6$\AA & 2.7\% & 2.3\% & 3.6\% & 6.5\% \\
H$\alpha -0.4$\AA & 5.9\% & 3.6\% & 5.1\% & 6.7\% \\
H$\alpha -0.2$\AA & 4.6\% & 3.8\% & 4.8\% & 7.7\% \\
H$\alpha \pm0.0$\AA & 3.4\% & 3.3\% & 4.4\% & 6.6\% \\
H$\alpha +0.2$\AA & 4.9\% & 4.3\% & 3.8\% & 5.6\% \\
H$\alpha +0.4$\AA & 6.3\% & 4.0\% & 5.6\% & 6.8\% \\
H$\alpha +0.6$\AA & 5.3\% & 4.2\% & 4.5\% & 7.3\% \\
%H$\alpha +0.8$\AA & 1.4\% & 3.3\% & 2.2\% & 4.9\% \\
%H$\alpha +1.0$\AA & 1.4\% & 3.7\% & 1.4\% & 3.2\% \\
H$\alpha +1.2$\AA & 1.0\% & 2.8\% & 1.5\% & 3.2\% \\
%H$\alpha +1.4$\AA & 0.9\% & 2.4\% & 1.1\% & 2.9\% \\
\hline
\end{tabular}
\end{table}
\subsubsection{Calcium 8542\AA\thinspace preferred models}
Before the flare onset in the 8542\AA\thinspace line, M2 fits are found concentrated over the sunspot umbra at the line core, with the results at $\pm$ 0.2\AA\thinspace looking similar (Figure~\ref{fig:preferred_8542_pre}). Moving away from the core, at the $\pm$0.4\AA\thinspace positions there are some M2 fits over the umbra but these are much more concentrated spatially. Further from the core, M2 fits are very scarce, with the majority of the pixels being described best by the standard coloured noise background (M1).

After the impulsive phase of the flare, the appearance of the 8542\AA\thinspace results is quite different (Figure~\ref{fig:preferred_8542_post}). There is an increased number of M3 fits, in particular in the line wings, and a decreased number of M2 fits in general. The -0.2\AA\thinspace position seems to show an enhancement of M2 fits over the umbra, while at the line core and at the +0.2\AA\thinspace position the number of M2 fits has dropped, and there is a change in the location of the oscillatory signals -- the positions where M2 fits occur is in fact quite different from the pre-flare case. The places showing M2 fits in the line core and +0.2\AA\thinspace positions during the post-impulsive period are now mostly at the northern umbra/penumbra border, rather than more towards the centre of the sunspot (compare with  Figure~\ref{fig:preferred_8542_pre}).

\subsubsection{H$\alpha$ preferred models}
The results for H$\alpha$ are in some ways very similar to the Ca \textsc{II} 8542\AA\thinspace case. For example, there are more M3 fits towards the far wings and in the post-impulsive data, however, unlike Ca \textsc{ii} 8542\AA\thinspace the M2 model fits are most prevalent in the +0.4\AA\thinspace and +0.6\AA\thinspace line positions, as opposed to the line core. Also, the area where M2 was the preferred fit is far smaller in both the pre-flare and post-impulsive cases than their counterparts in Ca \textsc{II} 8542\AA. Unlike for Ca \textsc{II} 8542\AA, there seems to be very little evidence of concentrated oscillatory signals before the flare occurs, even at the centre of the sunspot, except in the +0.4\AA\thinspace and +0.6\AA\thinspace panels of Figure~\ref{fig:preferred_Halpha_pre}.

After the impulsive phase, H$\alpha$ shows a large number of M3 fits which are not only confined to the wings of the line. Unlike the Ca \textsc{ii} 8542\AA\thinspace line, in many wavelength positions it appears that after the impulsive phase there has been an increase in the number of macropixels showing oscillatory signals, and we see lots of M2 fits occurring from the core out to $\pm$ 0.6\AA. 
%\lf{[Again I find this a bit hard to see by just eyeballing the images - e.g. it looks like the number of pink pixels at -0.4\AA has decreased in postflare compared to pre-flare. Could we somehow compare the pre/post flare number of model fits of each kind in a table or histogram?] } 
While the total area of M2 fits is fairly sparse compared to 8542\AA\thinspace, there is a considerable amount of oscillatory signal focused on the northern umbra/penumbra border where there was a lot less in the pre-flare results. 

\subsection{Periods and locations of Gaussian bump peaks}\label{sec:bump_peaks}
\begin{figure*}
    \centering
	\includegraphics[width=2\columnwidth]{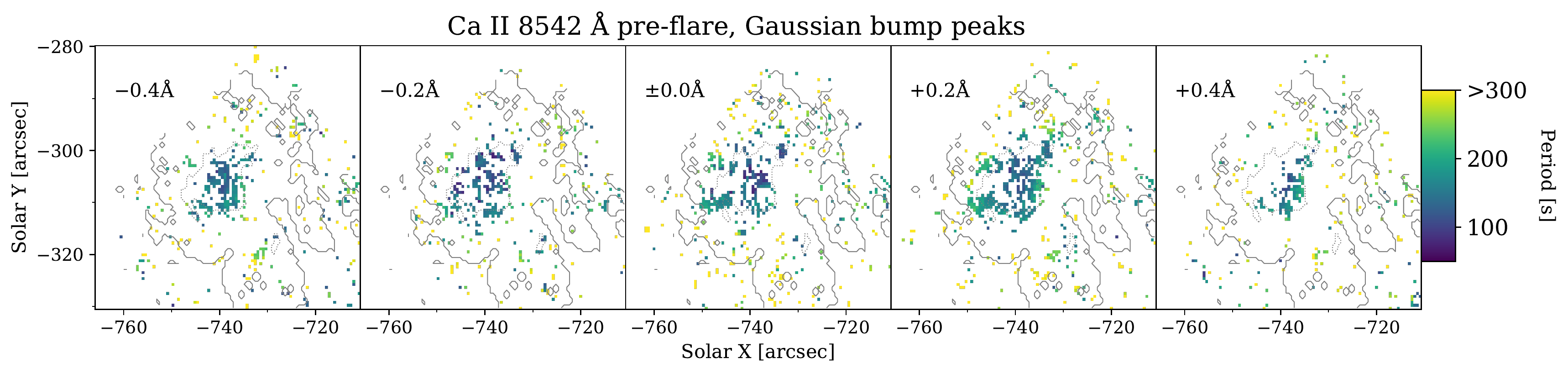}
    \caption{This plot is similar in layout to Figure~\ref{fig:preferred_8542_pre} but it shows periods at the peaks of Gaussian bumps, in macropixels where M2 was the preferred fit. Those macropixels which were not best fitted by M2 are left blank. This shows the difference in the period of oscillatory signals which are seen at different spatial positions across the images. In general, the period of the oscillations is shorter towards the centre of the sunspot.}
    \label{fig:peaks_ca8542_pre}
\end{figure*}

\begin{figure*}
    \centering
	\includegraphics[width=2\columnwidth]{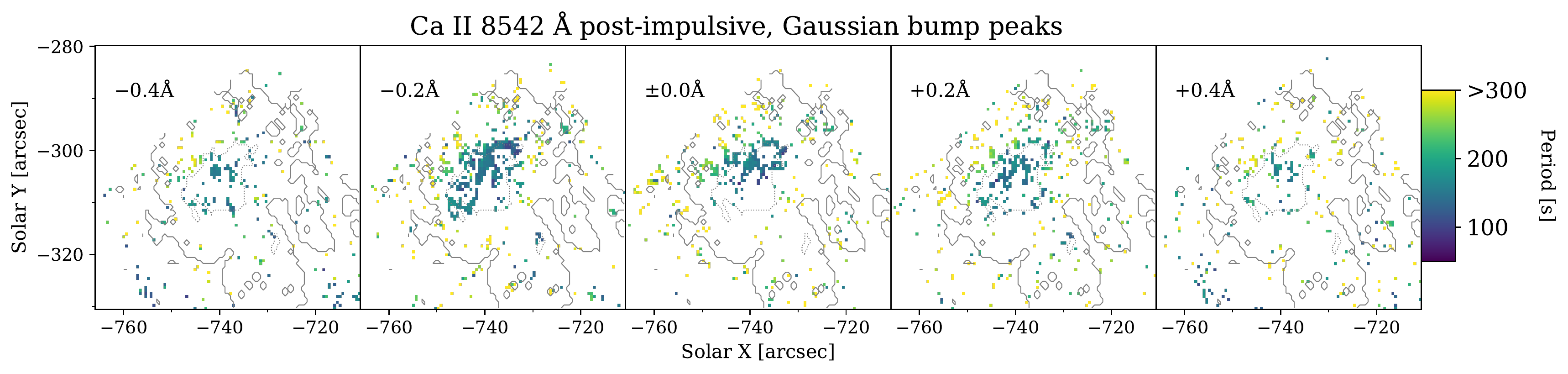}
    \caption{Similar to Figure~\ref{fig:peaks_ca8542_pre} but for the post-impulsive period. }
    \label{fig:peaks_ca8542_post}
\end{figure*}

\begin{figure*}
    \centering
	\includegraphics[width=2\columnwidth]{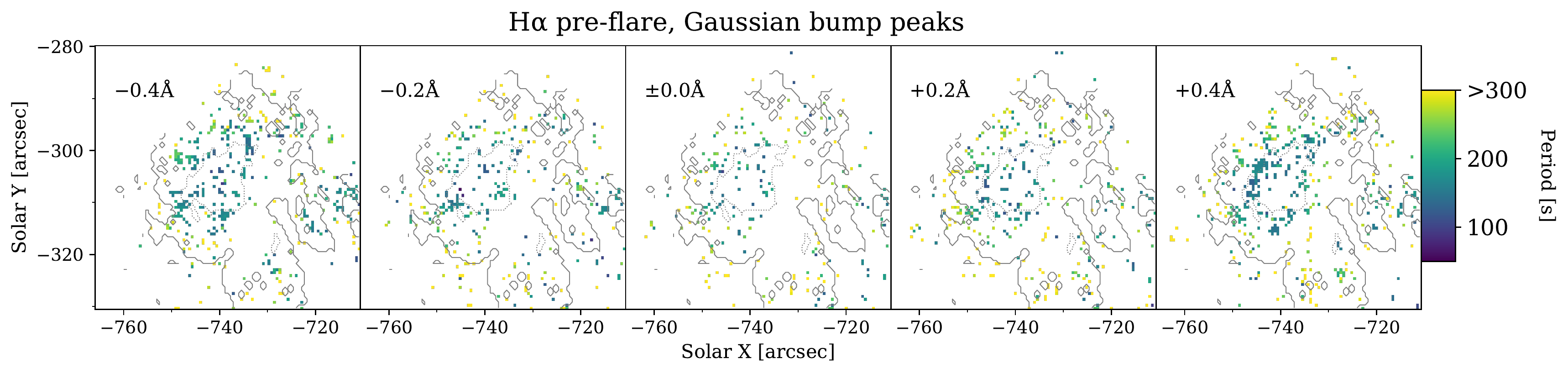}
    \caption{Similar to Figure~\ref{fig:peaks_ca8542_pre} but for the H$\alpha$ line.}
    \label{fig:peaks_Halpha_pre}
\end{figure*}

\begin{figure*}
    \centering
	\includegraphics[width=2\columnwidth]{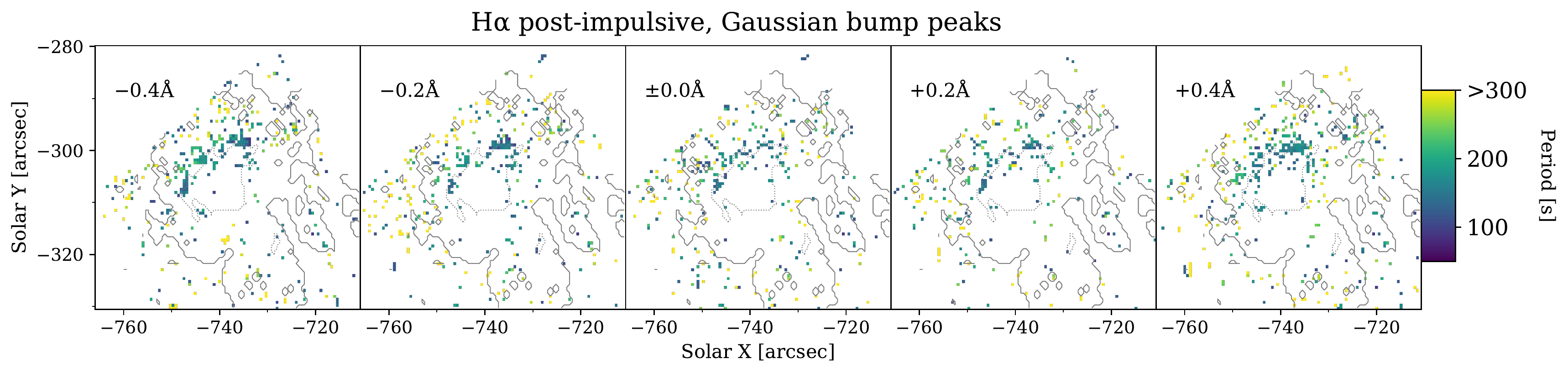}
    \caption{Similar to Figure~\ref{fig:peaks_ca8542_pre} but for H$\alpha$ in the post-impulsive period.}
    \label{fig:peaks_Halpha_post}
\end{figure*}

In pixels where M2 was the preferred fit, we can use the periods at which the Gaussian bumps in M2 peak to diagnose the characteristics of the oscillatory signals coming from those pixels. These results are shown in Figures~\ref{fig:peaks_ca8542_pre}-\ref{fig:peaks_Halpha_post} for the Ca \textsc{ii} 8542\AA\thinspace and H$\alpha$ lines, in both the pre-flare and post-impulsive time periods.

From these plots we see a difference in the periods of the bump peaks when moving from the centre of the sunspot outwards. In general the periods of the oscillatory signals are shorter in the centre of the sunspot, at 100-150s, and gradually increase into the 200-250s range towards the edges of the umbra. In the penumbra the periods reach 300s and above. This is most clearly seen in the Ca \textsc{ii} 8542\AA\thinspace data, but the effect is present in both lines.

These results could be interpreted similarly to those of \cite{Reznikova_2012_3mins} and \cite{Jess_2013ApJ...779..168J}, as evidence of a changing magnetic field inclination from the centre of the sunspot outwards.

There are some differences between pre-flare and post-impulsive results, with periods in post-impulsive data tending to be longer than in the pre-flare case. %\lf{[again a bit hard to see, histogram of number of pixels against period could help?]}.
Again this is clearer to see in Ca 8542\AA\thinspace results (compare the $\pm$0.2\AA\space and line core panels of Figures~\ref{fig:peaks_ca8542_pre} and \ref{fig:peaks_ca8542_post}), mostly due to the larger area of M2 pixels in this line, compared to H$\alpha$. This again could be caused by a differing magnetic field inclination. However, in this case the difference between pre-flare and post-impulsive period distributions could suggest that the magnetic field through the sunspot chromosphere has been affected by the flare itself, perhaps because of the reconfiguration of magnetic field taking place during the flare. 

\subsection{AIA Results}\label{sec:aia_results}
\begin{figure*}
    \centering
	\includegraphics[width=1.5\columnwidth]{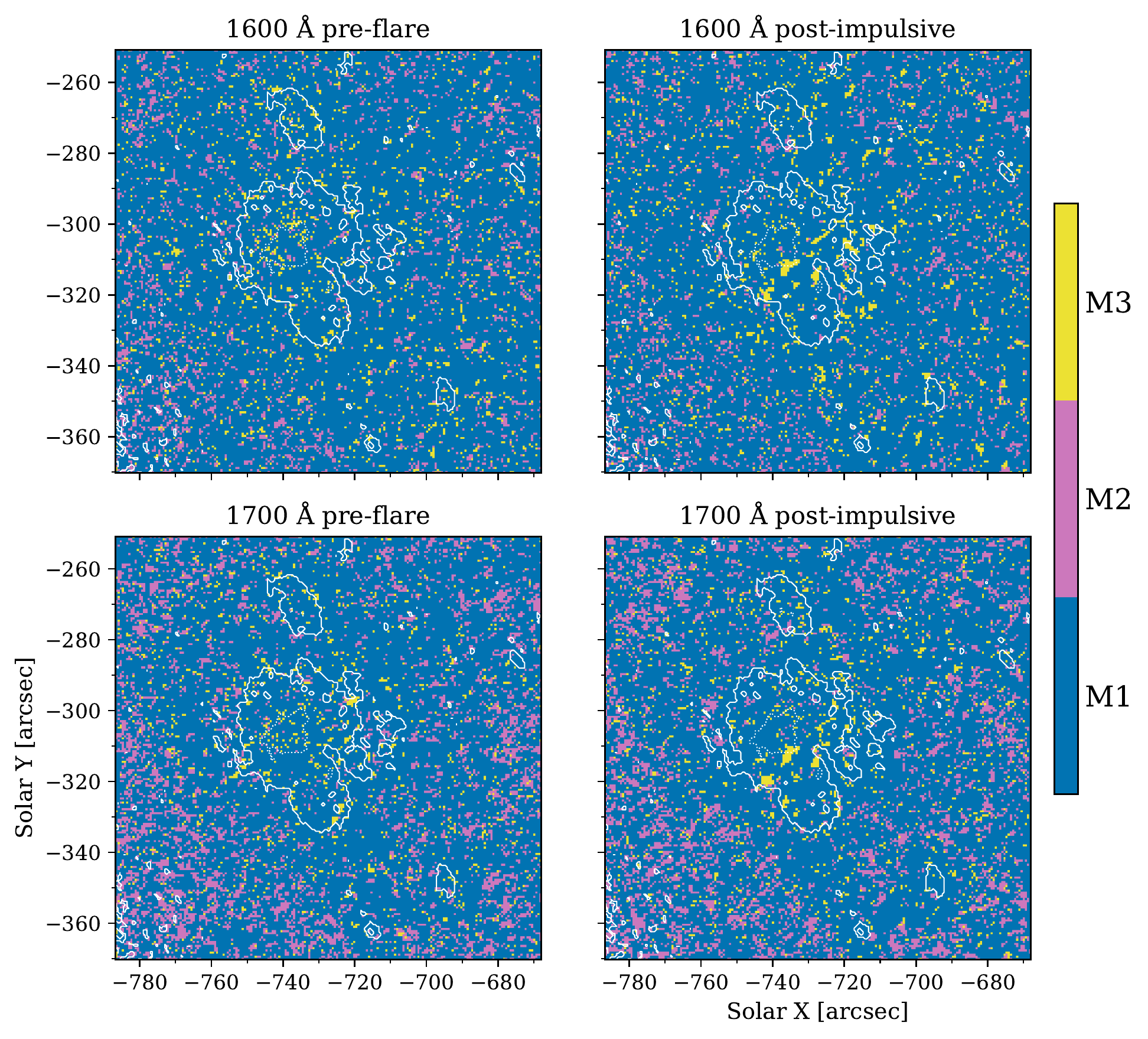}
    \caption{Similar to the corresponding plots for CRISP wavelengths (Figures~\ref{fig:preferred_8542_pre}-\ref{fig:preferred_Halpha_post}) but for the two AIA channels, with pre-flare and post-impulsive displayed on the left and right respectively. Note the field of view is slightly larger than in the CRISP figures. Overplotted are contours of 30\% and 73\% intensity from the first HMI continuum image to show the positions of the sunspot umbra and penumbra.}
    \label{fig:aia_prefer}
\end{figure*}

The preferred models found for pixels from AIA 1600\AA\thinspace and 1700\thinspace\AA\space are shown in Figure~\ref{fig:aia_prefer}, containing both pre-flare and post-impulsive results. These plots are in contrast to those in Figures~\ref{fig:preferred_8542_pre}-\ref{fig:preferred_Halpha_post} in that there is little convincing evidence whatsoever of concentrated oscillatory signals above the sunspot, or in the active region. Instead, there is an abundance of M2 fits seen outside the active region, around the edges of the plots. These match spatially with the intensity of the ultraviolet channels: immediately surrounding the sunspot umbra and penumbra is a large region of brighter plage, where scarce oscillatory signals are seen.

Table~\ref{tab:aia_stats} shows the percentage of M2 and M3 fits for the AIA channels. This table and also Figure~\ref{fig:aia_prefer} show that the 1700\AA\thinspace channel has far more pixels containing oscillatory signals than the 1600\AA\thinspace filter. There is also little change in the number of M2 fits between the pre-flare and post-impulsive periods in either channel.

%\lf{[Are there more or fewer M2 fits after the flare, or no difference? I am wondering whether we should compare with Ryan's findings of an enhanced oscillatory signal in 1600 and 1700. Though that was mostly during the flare I think which is the bit that we miss out, so it's not a direct comparison.]}

The most noticeable change between the pre-flare and post-impulsive results is the appearance in post-impulsive data of many locations best fitted by M3, at locations near the flare ribbons. As discussed in Section~\ref{sec:spectrum-models}, a successful fit to M3 does not directly tell us anything about oscillatory behaviour, however we can say that these timeseries must have deviated significantly from the coloured noise background, which is perhaps to be expected due to the large variations in brightness in these pixels.

\begin{table}
\centering
\caption{The percentage of all pixels in AIA wavelengths which had preferred fits M2 and M3, for the pre-flare and post-impulsive time periods.}
\label{tab:aia_stats}
\begin{tabular}{ccccc}
\hline
Wavelength & Pre M2 & Pre M3 & Post M2 & Post M3 \\
\hline
AIA 1600 \AA & 9.2\% & 4.9\% & 8.6\% & 5.7\% \\
AIA 1700 \AA & 15.8\% & 4.4\% & 15.9\% & 5.1\% \\
\hline
\end{tabular}
\end{table}

% \begin{figure*}
%     \centering
% 	\includegraphics[width=1.8\columnwidth]{0501_aia_peaks.pdf}
%     \caption{Similar to the corresponding plots for CRISP wavelengths (Figures~\ref{fig:preferred_8542_pre}--\ref{fig:preferred_Halpha_post}) but for the two AIA channels, with pre- and post-flare displayed on the left and right respectively. Overplotted are contours of 1600 \AA intensity to show the position of the sunpot umbra and penumbra.}
%     \label{fig:aia_peaks}
% \end{figure*}

\section{Discussion}\label{sec:discussion}
\subsection{Limitations}
While the spectral fitting method used here is a powerful technique and has been proven useful in many previous studies, there are limitations to its effectiveness.

One of the first things to consider is our choice of models. The results of this methodology are dictated by the models, as the data are forced to fit one of three models we choose. For example, if some process exists which produces two separate periods in the same timeseries, the resulting PSD could contain two bumps. This could either cause the curve fit algorithm to select one of these two bumps and not the other, or it could fail to fit the single bump model altogether. The case of double periods has been known to happen in QPP data \citep{Inglis_2009_multi-period}, and could be better handled by e.g. wavelet analysis. However wavelet analysis is difficult to automate for large datasets like the one studied here.

One of the main reasons the kappa-shaped model was fitted in addition to the coloured noise background and Gaussian bump models was due to the initial choice of models being insufficient. On inspection of M2 fit results by eye, some spectra were selected as an M2 fit when they had flattened off in the low frequency regime, instead of returning to the red noise background level. These poor results at the low frequency part of the PSDs could have been because the fitting technique is biased towards the high frequency end, due to the nature of the frequency space covered by the Fourier transform (more frequency points at high frequencies leads to a greater influence on the goodness of fit).

The CRISP observation for this dataset finished at approximately 17:30, putting limits on the length of the post-impulsive timeseries we can analyse. The length of the timeseries relates to the lowest frequency which can be studied using the Fourier transform. Longer timeseries would also reduce noise in the spectra we obtain, and hence lead to better (or at least, faster converging) fits. 

\subsection{Interpretations}
The CRISP results showed considerable variation across the wavelength steps of the spectral lines. This could be because different points in the lines sample different heights in the atmosphere: both H$\alpha$ and  Ca \textsc{II} 8542\AA\thinspace sample the mid-chromosphere at their cores \citep[$\sim$1Mm,][]{kuridze_mathioudakis_simoes_2015} and the upper photosphere at their wings. Strong, isolated oscillatory signals could  be present only above a certain height, beyond the altitude at which long period signals from the photosphere have decayed (as described in section~\ref{sec:intro}). Another effect which applies here is that radiation at a particular wavelength does not always originate from exactly the same height, as the contribution function can be somewhat spread out. An example of this can be seen in \cite{kuridze_mathioudakis_simoes_2015}, who used RADYN to replicate observed line profiles from the same flare as is studied in this paper. If there are local pressure disturbances in the atmosphere caused by MAG waves, then sampling from a wide range of heights would lead to muddied signals, with the potential for destructive interference occurring. MAG waves could be travelling through the lower parts of the sunspot chromosphere, but not be detectable in this kind of observation because of the broad contribution function. Conversely, if we receive light which is emitted from a more vertically compact region, it is more likely that signals will be preserved.

These reasons could also explain the results from the AIA UV channels in Section~\ref{sec:aia_results}. These channels are very broad filters and the heights which they sample are perhaps not precise enough to detect localised oscillations in this analysis. A study by \cite{Simoes_2019_AIA-channels} showed how the temperatures sampled by these filters can be different in flare and plage data. The results for the AIA 1600\AA\thinspace and 1700\AA\thinspace channels showed a lack of oscillatory signatures over the umbra, and more in the penumbra and plage. These results are similar to those obtained by \cite{Battams_2019_EUV-powerlaws} who used a power spectrum fitting analysis and found spectral bumps across almost the whole disk in these AIA channels, except in the area immediately surrounding a sunspot. A similar phenomenon named ``height inversion'' has been observed by other authors, where 3-minute signals are seen to be strong in chromospheric sunspot umbrae, but almost nonexistent in the photosphere \citep{Kobanov_2008AstL...34..133K, Kobanov_2011_A&A...525A..41K}. While \cite{milligan_2017} did find flare-related oscillatory signatures in 1600\AA\thinspace and 1700\AA\thinspace data, these were signatures integrated over a large field-of-view, and the flare-related oscillatory signals were during the impulsive phase which we cannot study due to saturation.

The positions where significant oscillatory bumps were identified in CRISP data was changed by the flare activity. Comparing the bottom right panel of Figure~\ref{fig:context} to the results in Figures~\ref{fig:preferred_8542_pre}-\ref{fig:peaks_Halpha_post}, we see that one of the flare ribbons develops over the lower corner of the sunspot umbra. This could explain the lack of M2 fits in this area in the post-impulsive results. The number of macropixels with M2 fits increased at the northern boundary of the umbra in most wavelengths at $\pm0.4$\thinspace\AA, and it is possible the same would have happened at the southern boundary but the natural oscillations, which are connected to the temperature of the plasma, may have been dramatically affected by flare heating.
 %this activity has been interrupted by \lf{[or affected by?]} the intense heating from the flare. The
If we assume the oscillatory signals we detect here are caused by MAG waves travelling from below, along  the sunspot field, the results from Sections~\ref{sec:preferred} and \ref{sec:bump_peaks} may be linked to properties of the magnetic field in the chromosphere above this sunspot. For instance, the areas which produced a lot of concentrated M2 fits are different in much of the pre-flare and post-impulsive results. This could be caused by the magnetic fields having a different orientation after the flare process, causing the MAG waves travelling along the field lines to be guided to a different spatial location. As an example, if we consider the chromosphere to be 2000km thick, and observe the locations of signals to move by $\sim$5 arcsec, this would correspond to a magnetic field line which was originally normal to the solar surface inclining by $\sim60$ degrees. Strong, flare-related changes in field inclination have previously been inferred from line-of-sight magnetograms \citep{Sudol_and_Harvey_2005ApJ...635..647S} or observed in vector magnetograms \citep{Petrie_2019ApJS..240...11P}.

%\lf{[Is it plausible that this is due to the same field becoming more tilted? The chromosphere is only ~2000km thick (=3 arcsec) but the locations are shifting by maybe 5-10 arcsec - or more? - meaning that there would have to be a very big change in field line angle. Or could it be that all the field gets tilted and the oscillatory signals are now seen on the fields with the `right' tilt, because their own acoustic cutoff frequency changes? But then it's not consistent with your discussion below, as the `right' tilt for detection of oscillations post-flare would presumably be the same tilt as pre-flare so the period would be the same. Let's discuss. Also we need to make the point clearly somewhere (if we don't already) that the oscillatory signals are not at locations of flare heating so we don't think that there are effects due to a flare-induced change of opacity.]}

Further evidence for this interpretation can be seen in the results of Section~\ref{sec:bump_peaks} where the periods at which Gaussian bumps peak changed after the flare event. In MAG waves the acoustic cut-off frequency determines the period of the oscillations:
\begin{equation}
    \omega_c = \frac{\gamma g\cos{\theta}}{2c_s} %= %\sqrt{\frac{\gamma \mu g^2}{4RT}}
    \propto \frac{g\cos{\theta}}{\sqrt{T}}
\end{equation}

 $\theta$ is inclination angle from vertical. Following this equation, the fact that $g$ has negligible variation, and observing that temperature change is unlikely to be a factor as the areas of M2 fits were far removed from the flare ribbons and do not show any intensity variations, the inclination angle is the only factor which could cause the change in cut-off frequency in this case. Following our example from above of a 60 degree inclination change, the cut-off frequency would be halved, and the prominent period would be doubled.

\begin{figure}
    \centering
	\includegraphics[width=1\columnwidth]{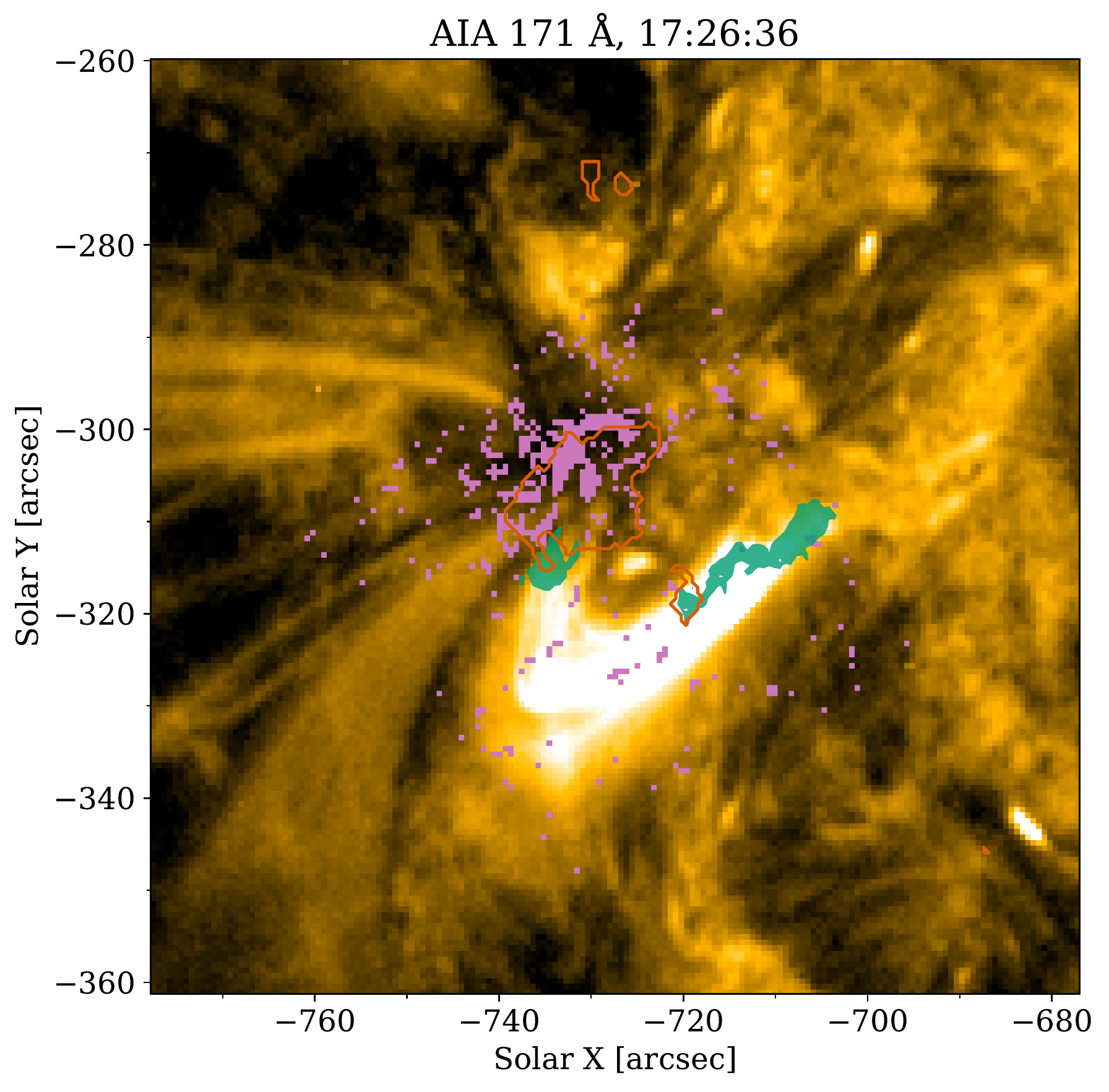}
    \caption{An image from the 171\AA\space passband from AIA at the end of the post-impulsive time period. Plotted in pink are the locations of macropixels with spectra which produced Gaussian bumps in the post-impulsive Ca \textsc{ii} 8542 $-0.2$\AA\thinspace results. The contours outline the sunspot umbra, using the 40\% intensity level of HMI continuum, and the solid patches show the flare ribbons, determined by 70\% of the maximum intensity value of 8542\AA\thinspace core at this time.}
    \label{fig:combine}
\end{figure}

To investigate our interpretations further, we have plotted in Figure~\ref{fig:combine} an image from AIA's 171\thinspace\AA\space channel, taken at the end of the CRISP observing window. This channel is the most suitable to get an impression of the magnetic activity in the active region. The flare ribbons seen in the CRISP line cores are linked by a newly formed hot loop, which develops after the flare activity into the bright structure shown in Figure~\ref{fig:combine}. Other larger loop structures are seen emerging from the active region, to the east and north-east of the flare footpoints and sunspot umbra. It can be seen that these loops undergo changes during the flare activity, with some contraction of the loops visible. It is clear from observing this particular wavelength that the magnetic structures in this active region have been altered during the flare activity and this gives weight to our interpretation regarding a change of the magnetic field inclination.

\section{Conclusions}\label{sec:conclusions}
We studied this active region to try and understand the types of oscillatory signals which can be affected or induced by flare activity. We found significant oscillatory behaviour consistent with the theory of MAG waves travelling up strong magnetic fields in sunspots, both before and after the flare. In this first spectrally-resolved analysis of flare chromospheric oscillations using the spectral fitting method, we found that the periodic signals seen in H$\alpha$ and Ca \textsc{ii} 8542\AA\thinspace line core observations were not seen in the line wings or the AIA ultraviolet channels, likely due to the broader range of heights sampled by these observations. 

There is evidence of the oscillatory behaviour being altered indirectly by the flare, both in the locations of the signals and the periods of the oscillations. The signals were found to have moved from covering almost the whole sunspot umbra before the flare, to being concentrated on the northern umbral border afterwards. There was a lack of signals at the locations of the chromospheric flare ribbons, most likely due to the intense heating of the plasma at these locations. In both pre-flare and post-impulsive results, the periods at which the oscillations were observed increases radially out from the umbra, but in post-impulsive data the periods are in general longer.

We believe these results are evidence of a changed magnetic environment in the sunspot as a result of the flare activity, and this interpretation is backed up by images of coronal loops which are connected to the site. Our work provides evidence of the ways solar flares can affect the solar atmosphere, in particular the chromosphere, and that the flare's influence can be felt over the whole active region.  

\section*{Acknowledgements}
DCLM would like to thank Aaron Reid and James Threlfall for insightful conversations, and the Carnegie Trust for the Universities of Scotland for support through PhD scholarship PHD007733. LF is grateful for support from UK Research and Innovation's Science and Technology Facilities Council under grant award numbers ST/P000533/1 and ST/T000422/1. The research leading to these results has received funding from the European Community's Seventh Framework Programme (FP7/2007-2013) under grant agreement no. 606862 (F-CHROMA). ROM would like to thank the Science and Technologies Facilities Council (UK) for the award of an Ernest Rutherford Fellowship (ST/N004981/1).

\section*{Data availability statement}
The CRISP data which were analysed in this study are available on the F-CHROMA database (\url{f-chroma.org}). The SDO data can be accessed from the Joint Science Operations Centre (\url{jsoc.stanford.edu}). Scripts used to generate these results and the figures in this paper can be found at \url{github.com/davidclmillar}.

%%%%%%%%%%%%%%%%%%%%%%%%%%%%%%%%%%%%%%%%%%%%%%%%%%



\bibliographystyle{mnras}
\bibliography{mnras_paper} % if your bibtex file is called example.bib

%%%%%%%%%%%%%%%%%%%%%%%%%%%%%%%%%%%%%%%%%%%%%%%%%%

%%%%%%%%%%%%%%%%% APPENDICES %%%%%%%%%%%%%%%%%%%%%

% \appendix

% \section{Some extra material}

% If you want to present additional material which would interrupt the flow of the main paper,
% it can be placed in an Appendix which appears after the list of references.

%%%%%%%%%%%%%%%%%%%%%%%%%%%%%%%%%%%%%%%%%%%%%%%%%%

% Don't change these lines
\bsp	% typesetting comment
\label{lastpage}
\end{document}